\documentclass[sigconf]{acmart}

\usepackage{algorithm}
\usepackage{algorithmicx}

\usepackage{adjustbox}
\usepackage[dvipsnames]{xcolor}

\usepackage{xcolor}
\definecolor{gray1}{HTML}{e4e4e4}
\definecolor{blue1}{HTML}{e0ffff}

\usepackage{amsmath}
\usepackage{amsfonts}
\usepackage{amsthm}

\usepackage{enumitem}
\usepackage{graphicx}
\usepackage{xspace}

\theoremstyle{definition}

\usepackage{listings}
\usepackage{textcomp}
\definecolor{prismgreen}{rgb}{0, 0.6, 0}
\newcommand{\prismfont}{\fontsize{6pt}{6.9pt}\selectfont}

\lstdefinelanguage{Prism}{ 
basicstyle=\color{red}\prismfont\sffamily, 
keywords={bool,C,ceil,const,ctmc,double,dtmc,endinit,endmodule,endrewards,endsystem,F,false,floor,formula,G,global,I,init,int,label,max,mdp,min,module,nondeterministic,P,Pmin,Pmax,prob,probabilistic,R,rate,Rmin,Rmax,S,stochastic,system,true,U,X,observables,endobservables},
keywordstyle={\bfseries\color{black}},
numberstyle=\tiny\color{black},
comment=[l] {//}, morecomment=[s]{/*}{*/}, 
commentstyle= \color{prismgreen}, 
tabsize=4, 
captionpos=b, 
escapechar=@, 
literate=%
        {-}{{\textcolor{black}{$-$}}}{1}%
        {->}{{\textcolor{black}{$\rightarrow{}$}}}{2}%
        {0}{{\textcolor{blue}{0}}}{1}%
             {1}{{\textcolor{blue}{1}}}{1}%
             {p1}{{\textcolor{red}{p1}}}{1}%
             {2}{{\textcolor{blue}{2}}}{1}%
             {p2}{{\textcolor{red}{p2}}}{1}%
             {3}{{\textcolor{blue}{3}}}{1}%
             {4}{{\textcolor{blue}{4}}}{1}%
             {5}{{\textcolor{blue}{5}}}{1}%
             {6}{{\textcolor{blue}{6}}}{1}%
             {7}{{\textcolor{blue}{7}}}{1}%
             {8}{{\textcolor{blue}{8}}}{1}%
             {9}{{\textcolor{blue}{9}}}{1}%
             {.}{{\textcolor{blue}{.}}}{1}%
             {=}{{\textcolor{black}{=}}}{1}%
             {[}{{\textcolor{black}{[}}}{1}%
             {]}{{\textcolor{black}{]}}}{1}%
             {;}{{\textcolor{black}{;}}}{1}%
             {+}{{\textcolor{black}{+}}}{1}%
             {*}{{\textcolor{black}{*}}}{1}%
             {:}{{\textcolor{black}{:}}}{1}%
             {\&}{{\textcolor{black}{\&}}}{1}%
             {|}{{\textcolor{black}{|}}}{1}%
             {?}{{\textcolor{black}{?}}}{1}%
             {"}{{\textcolor{black}{"}}}{1}%
             {(}{{\textcolor{black}{(}}}{1}%
             {)}{{\textcolor{black}{)}}}{1}%
             {'}{{\textcolor{black}{'}}}{1}%
}

\newcommand{\block}[1]{\textcolor{blue}{#1}} 

\definecolor{burntorange}{rgb}{0.8, 0.33, 0.0}
\newcommand{\revision}[1]{\textcolor{black}{#1}}

\AtBeginDocument{
  }

\copyrightyear{2026}
\acmYear{2026}
\setcopyright{cc}
\setcctype{by}
\acmConference[SEAMS '26]{21st International Conference on Software Engineering for Adaptive and Self-Managing Systems}{April 13--14, 2026}{Rio de Janeiro, Brazil}
\acmBooktitle{21st International Conference on Software Engineering for Adaptive and Self-Managing Systems (SEAMS '26), April 13--14, 2026, Rio de Janeiro, Brazil}
\acmPrice{}
\acmDOI{10.1145/3788550.3794869}
\acmISBN{979-8-4007-2445-9/2026/04}

\newcommand{\linelabel}[1]{}

\begin{document}

\title{Formally Guaranteed Control Adaptation for ODD-Resilient Autonomous Systems}

\author{Gricel V\'{a}zquez}
\affiliation{%
  \institution{Department of Computer Science, University of York, UK}
  \city{}
  \country{}}
\email{gricel.vazquez@york.ac.uk}

\author{Calum Imrie}
\affiliation{%
  \institution{Department of Computer Science, University of York, UK}
  \city{}
  \country{}}
\email{calum.imrie@york.ac.uk}

\author{Sepeedeh Shahbeigi}
\affiliation{%
  \institution{Department of Computer Science, University of York, UK}
  \city{}
  \country{}}
\email{sepeedeh.shahbeigi@york.ac.uk}

\author{Nawshin Mannan Proma}
\affiliation{%
  \institution{Department of Computer Science, University of York, UK}
  \city{}
  \country{}}
\email{nawshinmannan.proma@york.ac.uk}

\author{Tian Gan}
\affiliation{%
  \institution{Department of Computer Science, University of York, UK}
  \city{}
  \country{}}
\email{tian.gan@york.ac.uk}

\author{Victoria J Hodge}
\affiliation{%
  \institution{Department of Computer Science, University of York, UK}
  \city{}
  \country{}}
\email{victoria.hodge@york.ac.uk}

\author{John Molloy}
\affiliation{%
  \institution{Department of Computer Science, University of York, UK}
  \city{}
  \country{}}
\email{john.molloy@york.ac.uk}

\author{Simos Gerasimou}
\affiliation{%
  \institution{Cyprus University of Technology}
  \city{Limassol}
  \country{Cyprus}}
\email{simos.gerasimou@cut.ac.cy}




\renewcommand{\shortauthors}{V{\'a}zquez et al.}

\begin{abstract}
    Ensuring reliable performance in situations outside the Operational Design Domain (ODD) remains a primary challenge in devising resilient autonomous systems. We explore this challenge by introducing an approach for adapting probabilistic system models to handle out-of-ODD scenarios while, in parallel, providing quantitative guarantees. Our approach dynamically extends the coverage of existing system situation capabilities, supporting the verification and adaptation of the system's behaviour under unanticipated situations. Preliminary results demonstrate that our approach effectively increases system reliability by adapting its behaviour and providing formal guarantees even under unforeseen out-of-ODD situations.

  
  
\end{abstract}



 \maketitle


\section{Introduction}

The increasing integration of autonomous systems in safety-critical domains, such as healthcare~\cite{rafiq2025symbolic} 
and maritime transportation~\cite{lee2025enhancing}, demands rigorous safety guarantees. A primary challenge in developing resilient autonomous systems is ensuring reliable performance in situations that extend beyond their predefined Operational Design Domain (ODD). At design time, the ODD defines the specific conditions under which a system is intended to function safely, often based on established standards~\cite{SAE_J3016_201806}. However, the complexity of real-world environments\revision{, evolution of the working environment and domain,} and emergent system behaviours mean that systems could encounter conditions outside their ODD \cite{hodge2025outofdistributiondetectionsafetyassurance}. 

\linelabel{intro:unknown-unknowns}\revision{Such out-of-ODD conditions will cause the system behaviour to become inherently uncertain, due to the ODD at design time no longer fully applicable.} While improving the underlying system could help mitigate risks discovered at runtime, such updates are often not feasible at runtime, as they may require unaffordable redesign or retraining. Moreover, the system might not be equipped to detect such out-of-ODD conditions. Consequently, it becomes necessary to adapt the system's controller to avoid critical situations where safety requirements are violated at runtime. This ensures that the system can continue to operate safely even under previously unseen, and potentially unsafe conditions.

\begin{figure*}[t]
    \centering
    \includegraphics[width=0.75\linewidth]{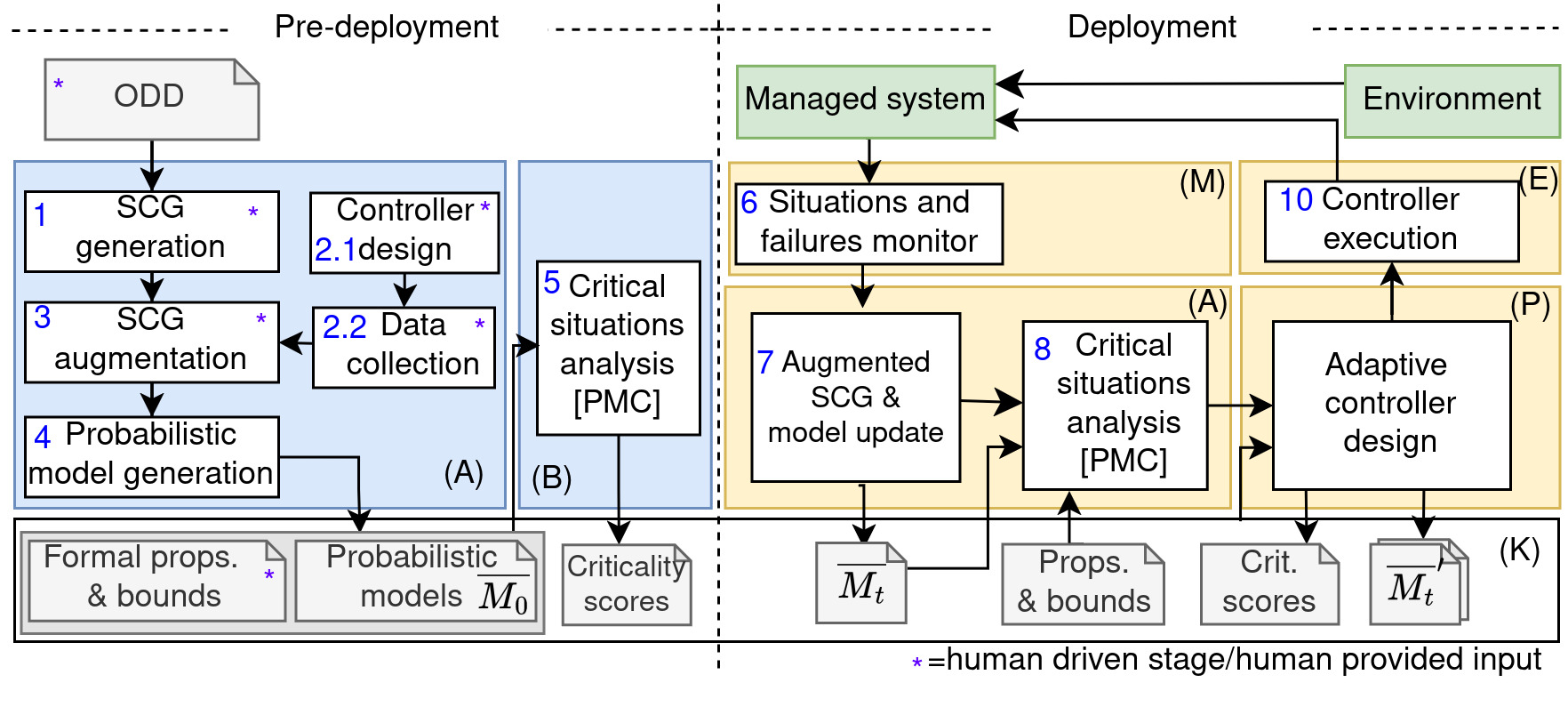}
    \vspace{-0.5cm}
    \caption{SAVE approach overview: pre-deployment (A,B) and deployment (M,A,P,E) phases with shared knowledge.}
    \label{fig:approach}
\end{figure*}

Self-adaptive Systems (SAS) identify changes and devise strategies to continue successful operation. Runtime approach-es (e.g., online testing and runtime verification) are valuable for SAS as they focus on observing the system's behaviour as it executes. While these methods are effective at detecting deviations from expected behaviour, they lack a framework for the mitigation of out-of-ODD encounters, and for dynamically adapting the system's underlying models to safely handle new situations while complying with strict safety requirements. To address these limitations, this paper introduces \textbf{SAVE} (\textbf{S}ituation-\textbf{A}ware \textbf{V}erification and control synth\textbf{E}sis), a\revision{n ODD-driven, situation-centric modelling and adaptation} approach, specifically designed to handle out-of-ODD conditions. Our approach leverages simulation to build probabilistic models of system behaviour, and probabilistic model checking to provide quantitative guarantees about the system's performance and safety. To ground our approach, we use an autonomous maritime system as a motivating example.

The main contributions of our paper are as follows:

\noindent
$\bullet$ A novel \linelabel{intro:novelty}\revision{situation-centric approach}, grounded in SAS principles leveraging
system situations extracted from the ODD \linelabel{intro:novelty2}\revision{to generate probabilistic situation models, and uses them to generate} controllers at deployment time, providing formal quantitative guarantees on safety and performance.

\noindent
$\bullet$ A method for encoding probabilistic transitions between encountered situations and a criticality score metric for the assessment of such situations, \linelabel{intro:novelty4}\revision{incorporating data-driven updates to transition probability values}. 

\noindent
$\bullet$ \linelabel{intro:novelty3}\revision{A runtime verification-driven adaptation loop that detects safety system violations at deployment time and synthesises updated controllers that remain safe (even under ODD drift).}

\vspace{-1.5mm}
\section{Background} 
\label{sec:background}

\vspace{1mm}\noindent
\textbf{Situation Coverage Grid (SCG)}. An SCG is a composite criterion used to test autonomous systems by systematically exploring a wide range of scenarios~ \cite{alexander2015situation, proma2025scaloft}. This grid combines various situational elements (e.g., road junction types and entities like cars, and pedestrians)~\cite{tahir2021intersection}.

\noindent
\textbf{Discrete-time Markov chain (DTMC).} A discrete-time Markov chain (DTMC) models a system that moves between states in discrete steps according to fixed probabilities~\cite{kwiatkowska2007stochastic}. 
Probabilistic Computation Tree Logic (PCTL) is a formal language for the specification of requirements~\cite{kwiatkowska2007stochastic, kwiatkowska2011prism}. 
Probabilistic model checking (\textbf{PMC}) is the process of automatically verifying whether the DTMC model satisfies such requirements~\cite{baier2008principles}. 
This automation is allowed by tools 
 such as PRISM~\cite{kwiatkowska2011prism}.



\section{The SAVE approach} 
\label{sec:methodology}



An overview of our SAVE approach is depicted in Figure~\ref{fig:approach} divided into pre-deployment and deployment phases.\footnote{In this section, the blue formatted text corresponds to the numbered SAVE stages in Figure~\ref{fig:approach}.}

\subsection{Pre-deployment}

\noindent
\textbf{SCG Augmentation (A).} The ODD provides a structured representation of the operating context. From this structured representation, a discrete set of situations that the system is expected to encounter is generated. In this paper, we define a situation $\rho$ as a $n$-tuple of valid subsets of (attribute:value) of the ODD: $\rho=(v_1,v_2,\ldots,v_n)$, where $v_i\in \text{ODD}$ for $i=1,\ldots,n$. SAVE \block{generates an SCG (1)} from such situations. \linelabel{approach:situationExample}\revision{As a motivating example in the maritime domain, an ODD consists of three attributes: density of detected vessels of type A, type B, and the shortest time to collision (TTC) with any neighbouring vessel. The first two can take values from ($none,low,high$), while the latter ($short, long$). A concrete situation is then defined as $\rho=(none,low,short)$.}


To capture the system’s dynamic behaviour as it transitions between different situations, SAVE extends the SCG by incorporating transition probabilities derived from \block{empirical test data (2.2)}. Stakeholders and domain experts can design and test various \block{controller designs (2.1)}. The objective during this process is to obtain system controllers that meet the pre-defined system requirements, while minimising the likelihood of entering critical (prone to requirement violations) situations during deployment. This data-driven stage results in an \block{augmented SCG~(3)}.

\begin{definition}[Augmented SCG]
    Let $\overline{\xi}=\{\xi_1,\xi_2,...\}$ be a set of failures. Given an SCG, $\overline{\rho}$, and a list of failures, $\overline{\xi}$, an augmented SCG is defined as a tuple $(S, \delta)$; where $S=\overline{\rho} \cup \overline{\xi}$, $\overline{\rho} \cap \overline{\xi}=\varnothing$ is a set of states; and the transition function $\delta : \overline{\rho} \to \mathit{Dist}(S)$ defines probabilistic transitions across all situations and failures.
    
\end{definition}

To analyse the system properties in \block{(4)}, SAVE generates a set of DTMCs $\overline{\mathcal{M}_0}$, each with a unique initial state representing one situation $\rho$, with transition probabilities derived from the augmented SCG. Each DTMC model has a set of states, where each state represents a situation or a failure, and encodes transitions between normal and failure states.
\revision{\begin{definition}[Probabilistic Model of $\rho_i$]\linelabel{approach:definition} Given an augmented SCG $(S, \delta)$ and a situation $\rho_i$, we define a DTMC $\mathcal{\overline{M}}[i] = (S', \bar{s}, \delta',$ $AP, L)$; where $S'$ denotes the set of states, consisting of the state variables from $S$ in the augmented SCG; $\bar{s}\in S'$ is initial state representing the system starting at situation $\rho_i$; $\delta'$ is the transition function $\delta':S' \rightarrow Dist(S')$, where transitions, representing a change in situation or into a failure state, are obtained from the augmented SCG (failure states are sink states); $AP$ is a set of atomic propositions defined for failure states; and $L$ the state labelling function (see Section~\ref{sec:background}).
\end{definition}}



\textbf{Critical Situations Analysis (B).} After eliciting the situation-aware probabilistic models, SAVE applies probabilistic model checking (PMC) to automatically identify and rank the most critical situations. For each situation $\rho_i$ with DTMC $\overline{\mathcal{M}}[i]$, the system verifies the model against the elicited safety properties $\overline{\Phi}[k]$. The degree of property violation determines the \block{criticality score (5)}, where a score of 0 indicates requirement compliance, and higher values indicate increasing deviation from the property bound. As an example, consider the requirement "the minimum probability of success is 0.96." Here, 0.96 serves as the bound. A situation's DTMC model that achieves only 0.85 for this property has a higher criticality score \revision{(0.11)} than a model achieving 0.94 \revision{(0.02), although both models violate such property}. At this stage, normalisation techniques can be applied to ensure fair comparison when properties use different scales (falling outside the scope of this paper). 

During pre-deployment, the resulting criticality scores enable stakeholders to prioritise mitigation efforts by identifying the most critical situations and leveraging detected requirement violations to iteratively refine the controllers and overall system design \revision{until all tested controllers are safe (by constraining or modifying the system's behaviour, degrading soft requirements, etc.)}. Once this phase is successfully completed, the system progresses to the deployment stage.

\vspace{2mm}
\subsection{Deployment}

At deployment, the system is continuously adapted to ensure safe operation. SAVE is aligned with the MAPE-K loop as shown in Figure~\ref{fig:approach}. We define each stage as follows.

\vspace{1mm}
\textbf{Monitoring of the Managed System (M). } The \block{monitoring~(6)} stage obtains the following data at time $t\in \mathbb{R}^{>0}$: 1)~\textbf{changes in the current system's situation}, which monitors the current encountered situation $\rho_{t}$; and 2)~\textbf{system failures}, the known failure conditions pre-identified as $\overline{\xi}$. \linelabel{appr:howChangesMonitored}\revision{For our maritime domain example, the vessel must be capable of detecting other vessels, classifying them as type~A or~B based on their characteristics, and estimating the distance to the nearest vessel in order to determine the current situation.}

\vspace{1mm}
\textbf{Model Update and Analysis of Critical Situations (A). } SAVE uses the monitored data to construct a new augmented SCG at time $t$ \linelabel{approach:modelsUpdate}\revision{updating the transition probabilities defined by $\delta$ using frequentist~\cite{alasmari2022quantitative,calinescu2015formal} or Bayesian-based~\cite{zhao2024bayesian,zhao2020interval} approaches as the system evolves from one situation into another, or into a failure state \textcolor{blue}{(7)}. A new set of models $\overline{\mathcal{M}_t}$ is then constructed as in the pre-deployment. SAVE then analyses \textcolor{blue}{(8)} these models and, where necessary, updates them to derive a safe system controller as follows:}

$\bullet$ SAVE obtains the model where the current situation $\rho_t$ is an initial state. This model is analysed using PMC to assigned a critically score as in the pre-deployment stage. 

$\bullet$ If no violations were detected (criticality score$\leq0$), the system proceeds as normal.

$\bullet$ Else, SAVE obtains the criticality score of each model in $\overline{\mathcal{M}}_t$. The situation $\rho$ with the model with the worst criticality score is then obtained. Finally, all outgoing transitions in $\mathcal{M}_{\rho_t}$ from the state representing $\rho$ are removed, leaving only a self-loop with probability~$1$. This also means that such outgoing transition probability in the augmented SCG were set to zero. SAVE systematically removes the most critical situations until no violations remain (or until a predefined maximum number of situation states become sink states, indicating failure to synthesise a safe controller). Thus, SAVE prevents the system from continuing once an out-of-ODD (potentially unsafe) situation is detected. 


\vspace{1mm}
\textbf{Controller Synthesis (P). } As the controllers are modelled to be safe from the pre-deployment stage, at runtime, such controllers might become unsafe, for example, when out-of-ODD situations appear. When the analysis stage (A) detects that the current system configuration $\overline{\mathcal{M}_t}$ violates one or more requirements $\overline{\Phi}$, SAVE triggers an \block{adaptive controller design (9)} process. The aim is to synthesise a verifiably safe controller by iteratively excluding the most critical situations from the operational context. In each iteration, the situation's model with the highest criticality score, indicating the largest deviation from a safety requirement, is removed. 

After a critical situation and its model are removed, stage (9) modifies the remaining probabilistic models by modelling that situation state as a "sink state", effectively building a barrier that prevents the system from continuing from those discovered, unsafe conditions. Since removing one situation and altering the models can change the probabilistic outcomes of the entire system, the criticality scores for all remaining situations are then re-calculated. This cycle of identifying the most critical situation, blocking access to it, and re-evaluating the system's safety continues until all situations comply with the set of requirements (i.e., criticality score of zero), or a maximum number of iterations is reached. The final output is a revised set of safe situations and models ($\overline{\rho}_t', \overline{\mathcal{M}}_t'$), which constitute the newly synthesised, safer controller.  

When a system is in a situation where multiple candidates controllers exists from the pre-deployment stage, the selection is based on two criteria: the controller must result in no property violations, and it must minimise the probability of the system reaching an out-of-ODD situation. If more than one comply with such criteria, one is chosen at random.

 \vspace{1mm}
\textbf{Controller Execution (E).}  Finally, the selected controller is \block{executed~(10)} by instrumenting the managed system. 
The entire execution process continues, ensuring that the system's adaptation remains responsive to changes.

\section{Preliminary Evaluation}
\label{sec:evaluation}

\textbf{Maritime domain case study.} We perform an initial evaluation of SAVE using a simplified maritime case study involving a Marine Autonomous Surface Ship (MASS) navigating among two types of crewed vessels (A and B). Each vessel type is characterised by its size, velocity, and time to collision (TTC) (short or long). The MASS, controlled by an adaptive AI-based controller, must avoid collisions and maintain adequate separation from other vessels. The ODD is deliberately simplified and discretised to keep the scenario tractable while preserving the essential dynamics of encounter situations such as head-on, crossing, and overtaking. 

There are two interlinked monitored failures: \textit{f1} corresponding to inadequate time to react to avoid collision when the TTC is too short; \textit{f2} signifying a a near catastrophic collision when the distance between vessels in near to zero units. In extreme scenarios such as when a vessel is travelling at a high speed, both might be detected at the same time. Finally, the following system properties are considered:

\begin{itemize}
    \item ($\Phi_1$) The probability of failure $f_1$ occurring within the next 50 situations must be less than $0.99$:
    $\mathrm{P}_{=?}\,[\,F^{\leq 50}\, f_1\,]< 0.99$.
    
    
    \item ($\Phi_2$) The probability of failure $f_2$ occurring within the next 50 situations must be less than $0.95$: $\mathrm{P}_{=?}\,[\,F^{\leq 50}\, f_2\,]< 0.95.$
 \end{itemize}

\begin{table}[t]
\centering
\caption{SAVE adaptation \revision{results after a violation is detected. Without adaptation, the baseline fails in all cases.}}
\resizebox{0.45\textwidth}{!}{
\begin{tabular}{lp{1.5cm}p{2.1cm}p{2.1cm}p{2.6cm}}
\hline
\begin{tabular}{@{}l@{}}\textbf{ID}\end{tabular} &
\begin{tabular}{@{}l@{}}\textbf{Property}\\ \textbf{violated}\end{tabular} &
\begin{tabular}{@{}l@{}}\textbf{Worst critic-}\\\textbf{ality score}\end{tabular} &
\begin{tabular}{@{}l@{}}\textbf{SAVE success}\\ \textbf{(no violations)}\end{tabular} &
\begin{tabular}{@{}l@{}}\textbf{Critical situations}\\ \textbf{avoided}\end{tabular} \\

\hline
1  & {[$\Phi_2$]}       & 0.03890 & True & {[s2, s3]} \\
2  & {[$\Phi_2$, $\Phi_1$]}   & 0.03482 & True & {[s2, s3, s5]} \\
3  & {[$\Phi_2$, $\Phi_1$]}   & 0.03556 & True & {[s3, s4]} \\
4  & {[$\Phi_2$, $\Phi_1$]}   & 0.04483 & True & {[s5, s1, s3, s2]} \\
5  & {[$\Phi_2$, $\Phi_1$]}   & 0.04905 & False & - \\
...16 & {[$\Phi_2$]}       & 0.04994 & True & {[s1]} \\
17 & {[$\Phi_2$]}       & 0.04999 & True & {[s3, s4]} \\
18 & {[$\Phi_1$]}       & 0.00998 & True & {[s3]} \\
19 & {[$\Phi_2$, $\Phi_1$]}   & 0.04999 & False & - \\
20 & {[$\Phi_2$, $\Phi_1$]}   & 0.04999 & False & - \\
\hline
\end{tabular}
}
\label{tab:rq1_results}
\end{table}

\vspace{1mm}
\noindent
\textbf{Research questions (RQs)}. We evaluate SAVE on 3 RQs.

\textbf{RQ1} [Effectiveness]. How effective is SAVE in reducing requirement violations caused by out-of-ODD situations?  

\textbf{RQ2} [Adaptation]. How effective is SAVE's adaptation in synthesising violation-free controllers compared to a baseline?

\textbf{RQ3} [Scalability]. Given that the most computationally expensive part of SAVE is obtaining the criticality score via PMC, how computationally effective is SAVE in synthesising new controllers as the number of situations increase?

\begin{table}[t]
\centering
\caption{SAVE adaptation example.}
\label{tab:rq3_adaptation}
\resizebox{0.49\textwidth}{!}{
\begin{tabular}{p{2.3cm}p{4.2cm}p{3.6cm}}
\toprule
\textbf{Event} & \textbf{Baseline Outcome} & \textbf{SAVE Response} \\ \midrule
Vessel deceleration (out-of-ODD) & Collision expected within the next 50 timesteps ($f2$), detected at time $t_1$. Collision happens in situation $i$ at time $t_2$. & Controller adapted at $t_1$. Critical situation $i$ and collision avoid. \\
Requirement $\Phi_1$ (safe TTC) & Violated at time $t_2$. & Satisfied post-adaptation at time $t_3$\\
Requirement $\Phi_2$ (near collision) & Violated at time $t_4$. & Satisfied post-adaptation at time $t_5$. \\
\bottomrule
\end{tabular}
}
\end{table}

\vspace{2mm}
\noindent
\textbf{Results and discussion}. 
We compare SAVE with a baseline system in which requirement violations may occur (assuming a fixed controller) to assess its effectiveness in reducing both requirement violations and collisions (\textbf{RQ1}). The number of situations to avoid is limited to four. For each run, vessel positions and speeds are randomly initialised to generate diverse situations and transition probabilities in the underlying probabilistic augmented SCG. Table~\ref{tab:rq1_results} reports \revision{10 out of 20 variants} where out-of-ODD perturbations (e.g., increased vessel velocity) lead to violations of one or both properties. In these experiments, we introduce randomly generated disturbances to the transition probability matrix to model such uncertainty. \linelabel{eval:GitHub}\revision{Code and complete experimental results are available in our \textbf{GitHub}~\cite{GitHub}}.

\linelabel{eval:PRISMautomated}\revision{We use PRISM~\cite{kwiatkowska2011prism} to obtain the criticality scores}. The results show that SAVE was able to avoid requirement violations in 14 of the 20 variants (column~4) by proactively avoiding high-risk situations (column~5). This improvement arises from SAVE’s ability to identify and adapt the system's controller to critical situations. In the maritime case study, this corresponds to performing a crash stop COLREGs, i.e., a collision avoidance manoeuvre when critical situations are detected.\footnote{The \textit{International Regulations for Preventing Collisions at Sea (COLREGs)}, published by the International Maritime Organization (IMO), provide guidance on how vessels should operate, including the rules governing typical encounters, such as head-on (rule 14); crossing (rule 15); and overtaking (rule 13).} All variants where SAVE failed to maintain a safe MASS controller involved violations of property $\Phi_2$, with their criticality score exceeding 0.04 (column~3). These insights from SAVE can support stakeholders in the re-design and refinement of autonomous control strategies.

For \textbf{RQ2}, Table~\ref{tab:rq3_adaptation} illustrates SAVE’s controller adaption strategy during a representative out-of-ODD event that causes a system violation, as well as individual system violations caused by the model updates as new data are available after deployment. In this scenario, the baseline system leads to a collision within the next 50 timesteps ($\Phi_2$), detected at time $t_1$ and occurring at time $t_2$. In contrast, SAVE identifies the event at $t_1$, computes the associated criticality score, and adapts the controller configuration to exclude the high-risk situation. This intervention prevents the collision entirely, demonstrating SAVE’s capacity to react to unforeseen dynamic changes in real time.

Continuing, SAVE restores compliance with both safety requirements. Requirement $\Phi_1$ (safe time-to-collision) and $\Phi_2$ (near-collision avoidance), which are violated in the baseline case at times $t_2$ and $t_4$, respectively, are both satisfied post-adaptation at times $t_3$ and $t_5$. These results confirm that SAVE not only mitigates imminent risks but also ensures sustained adherence to safety requirements under dynamic and uncertain conditions. 

\begin{figure}
   \centering
   \includegraphics[width=0.85\linewidth]{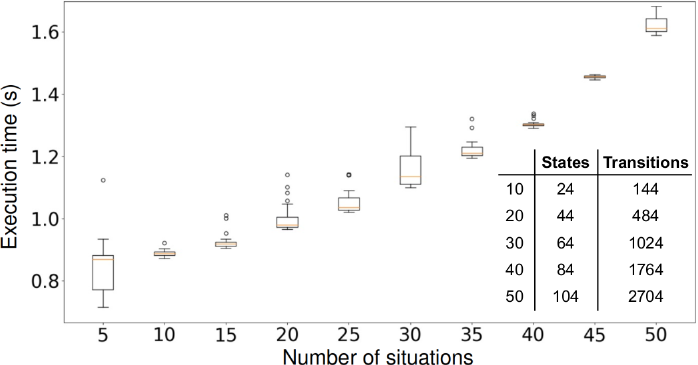}
   \caption{SAVE execution times to get criticality scores.}
   \label{fig:execution_times}
   \vspace{-7mm}
\end{figure}

Finally, as the most expensive part of SAVE is the generation of criticality scores using PMC, Figure~\ref{fig:execution_times} shows the scalability results (\textbf{RQ3}) by measuring the execution times for different numbers of situations. The time-scale in seconds show preliminary insights into the feasibility of SAVE for running verification at runtime for critical systems such as in the maritime domain.

\linelabel{eval:scalability}\revision{The results include the number of states and transitions as the number of situations increase. As expected, the state size increases linearly with the number of situations, and the number of transitions increases exponentially. However, every state in these models have transitions to every state with probability greater than 0. This can therefore be considered the worst case with regards to scaling, with some applications not having this aspect, resulting in far fewer additional transitions with increasing number of situations.} Overall, these preliminary results indicate that SAVE has the potential to effectively mitigate safety risks in dynamic environments, outperforming static controllers and demonstrating the feasibility of situation-aware, verifiably safe adaptation.

\section{Related Work}
\label{sec:relatedWork}

Adapting autonomous systems (AS) to safely handle unknown situations at runtime "\textit{is the ultimate challenge for self-adaptive systems}" \cite{cardozoD21seams}. Recent studies \cite{vazquez2025adaptive, purandare2023, avilés2024, imrie2024aloft,rafiq2025symbolic} present self-adaptive mechanisms enabling AS to adjust their planned paths and  system controllers under uncertainty. However, these approaches do not verify the autonomous systems (ASs) correctness at runtime. 

\revision{Runtime quantitative verification using PMC has been extensively studied as a means to provide formal guarantees for adaptive and self-adaptive systems (e.g.,~\cite{calinescu2018using,camara2015optimal,moreno2015proactive, calinescu2022discrete}). Existing approaches typically assume a given system model at runtime—such as Markov decision processes or stochastic games—and focus on synthesising adapted controllers. In contrast, SAVE adopts an ODD-grounded, situation-centric modelling approach in which probabilistic models are derived directly from semantically meaningful situations extracted from the system’s ODD. We also establishes explicit traceability between pre-deployment ODD and specifications, testing data, and runtime verification, an aspect not addressed by existing frameworks.}

Ideally, ASs learn and adapt over time, detecting and managing uncertainty~\cite{weyns2020introduction,weyns2023towards}, while being verified throughout their lifecycle to ensure acceptable safety \cite{hodge2025agile,rafiq2025symbolic}. A feedback loop is proposed in~\cite{abeywickrama2025} to integrate runtime verification insights and iteratively refine mission parameters, system architectures, and safety analyses. 
Out-of-distribution (OOD) detection identifies uncertainty in ASs which can lead to unpredictability \cite{hodge2025outofdistributiondetectionsafetyassurance,weyns2023self}. Simulation-based testing and formal methods can then be used to verify ASs \cite{torben2023formal}. 

In the maritime domain, ODDIT \cite{isaku2025digitaltwinbasedoutofdistributiondetection} uses digital simulation with ML models to assess if AS states are OOD, while \cite{gao2025FEL} propose an uncertainty-aware OOD detection method combining global and local trajectory models. Neither approach performs verification, and both rely on situation coverage (SitCov) testing \cite{nawshin2023,proma2025scaloft}, which grids the operational space to track tested conditions and quantify exposure to expected and novel situations. 
 
High-level decision-making and control of ASs often use finite-transition systems, suitable for verification via model checking \cite{torben2023formal,gerasimou2014efficient}. SitCov is used at design time for verification of situation models in~\cite{proma2025probabilistic}. The proposed SAVE approach builds on our previous work~\cite{proma2025probabilistic} by bridging the gap between formally verifying system compliance with safety properties and identifying violations of these requirements arising from deviations in the ODD relative to the expected values observed during pre-deployment testing. Wider adoption of such techniques depends on the development of comprehensive software frameworks, such as SAVE.

\section{Conclusions and Further Work}
\label{sec:conclusionsAndFurtherWork}


The paper introduces SAVE, a novel approach for adapting probabilistic models for autonomous systems to handle out-of-ODD situations that can impact system requirements, while providing quantitative guarantees. SAVE uses situation coverage and PMC to ensure that autonomous systems can comply with safety and reliability requirements, even when encountering novel or edge-case situations. Preliminary results show the feasibility of our approach to synthesise violation-free controllers at runtime.

Future work will incorporate our end-to-end simulated maritime example provided by our industrial partner.  We aim to implement comprehensive safety mechanisms that address potential failures, such as adaptive control and crash stop manoeuvre, in critical situations. Additionally, we will extend our evaluation on the adaptation of different control strategies; and explore techniques such as~\cite{calinescu2025ultimate} for the verification of large-scale systems. Finally, we will evaluate our approach across other cyber-physical system domains to provide deeper insights into the adaptability of our framework in diverse real-world scenarios.

\textbf{Acknowledgements.} This research was supported by the Centre for Assuring Autonomy (CfAA), a partnership between Lloyd’s Register Foundation and the University of York (\url{https://www.york.ac.uk/assuring-autonomy/}).

\newpage

\bibliographystyle{ACM-Reference-Format}
\bibliography{bibfile}


\end{document}